\documentclass[english]{article}
\usepackage[affil-it]{authblk}

\usepackage[T1]{fontenc}
\usepackage[latin9]{inputenc}
\usepackage{refstyle}
\usepackage{color,hyperref}
    \catcode`\_=11\relax
    \newcommand\email[1]{\_email #1\q_nil}
    \def\_email#1@#2\q_nil{%
      \href{mailto:#1@#2}{{\emailfont #1\emailampersat #2}}
    }
    \newcommand\emailfont{\sffamily}
    \newcommand\emailampersat{{\color{red}\small@}}
    \catcode`\_=8\relax

\makeatletter

\RS@ifundefined{subref}
  {\def\RSsubtxt{section~}\newref{sub}{name = \RSsubtxt}}
  {}
\RS@ifundefined{thmref}
  {\def\RSthmtxt{theorem~}\newref{thm}{name = \RSthmtxt}}
  
  {}
\RS@ifundefined{lemref}
  {\def\RSlemtxt{lemma~}\newref{lem}{name = \RSlemtxt}}
  {}

\usepackage{amsmath}

\makeatother

\usepackage{babel}
\usepackage{mathrsfs}
\begin{document}
\title{Unitary representation of the Poincar\'e group for classical relativistic dynamics }
\author{A. D. Berm\'udez Manjarres}
\affil{\footnotesize Universidad Distrital Francisco Jos\'e de Caldas\\ Cra 7 No. 40B-53, Bogot\'a, Colombia\\ \email{ad.bermudez168@uniandes.edu.co}}

\maketitle
\begin{abstract}
We give a unitary irreducible representation of the proper Poincar\'e
group that leads to an operational version of the classical relativistic
dynamics of a massive spinless particle. Unlike quantum mechanics,
in this operational theory there is no uncertainty principle between
position and momentum. It will be shown that the theory contains the
Koopman-von Neumann formalism as a particular case, and a explicit
connection with relativistic Hamiltonian mechanics will be given. 
\end{abstract}

\section{Introduction}

The structure of the Poincar\'e group and algebra plays a fundamental
role in the formulation of any relativistic theory. Quantum mechanics
and quantum field theory, being theories of operators acting on a
Hilbert space, lead to the study of unitary representations of the
Poincar\'e group. There is a vast literature on the subject, we refer
to the books \cite{weinberg,ohnuki} and references therein. On the
other hand, the theory of Poisson brackets representation of Lie groups
is the most natural way to study the Poincar\'e group in the context
of classical Hamiltonian mechanics \cite{prosperi,sudarshan}. 

Classical mechanics can be recast in the same mathematical language
of quantum mechanics. This old approach is due to Koopman \cite{cla1}
and von Neumann \cite{cla 2}. Whether for derivation of purely classical results or for comparison between quantum and classical mechanics, the  Koopman-von Neumann
formalism (hereafter abbreviated as KvN) has received increasing attention in the past two decades (see \cite{cla 3,cla 5,cla 6,Cla9,Cla 7,Cla 8,minimal coupling,cla9,cla10,cla11,cla12,cla13,cla14,cla15,cla16}), the possibility of formulating quantum-classical hybrid theories has also increased the interest in this formalism \cite{cla17,cla18,cla19,cla20}. The existence of the KvN theory raises the question of the
classification of the unitary representations of the groups of space-time
symmetries in the context of classical mechanics. An irreducible unitary
representation of the Galilei group that leads to classical mechanics
was given recently \cite{galilei}, and it was used to study the limitation
imposed by Galilean covariance on quantum-classical hybrids systems
\cite{galilei2}. 

In this paper we will give a new kind of unitary representation of
the proper Poincar\'e group. We do not attempt to give a complete classification
of this kind of representations, we prefer to focus on a single relevant
case. This classical representation differs from the quantum one in
the use of non-observable operators, what Sudarshan refers to as hidden
dynamical variables \cite{sudarshan-2}. The representation given
here results in the correct classical relativistic dynamics of a massive
spinless particle.

The KvN theory starts with the Liouville equation, and from it a Hilbert
space and a set of relevant operators are built. We could construct
a unitary representation of the Poincar\'e group starting directly from
the KvN theory, but we will take a different approach as we want to
avoid the use of previous results from analytical mechanics. Instead,
from the beginning we will postulate the existence of a Hilbert space
$\mathcal{H}_{C}$, and then we will look for a realization of the
proper Poincar\'e group where the space-time transformations are represented
by unitary transformations acting on $\mathcal{H}_{C}$. We will demand
the states belonging to $\mathcal{H}_{C}$ to contain all the information
that can classically be known about a particle, this leads to an absence
of uncertainty principles between the observables of the theory. We
will show that the theory constructed this way contains the KvN theory
as a particular case.

The structure and methods of this paper are very similar to the one
in \cite{galilei}, we just adapt them to the case of the Poincar\'e
group. This work is organized as follows: in the following section
we give a brief summary of the proper Poincar\'e algebra, and we present
the conventions that will be followed throughout this work. For reasons
of necessity, our notation for the elements of the Poincar\'e algebra
is not the standard one used in quantum mechanics.

In section 3 we introduce the Hilbert space $\mathcal{H}_{C}$ where
the operators from the Poincar\'e group act upon, and we state the action
of these operators over the vectors of $\mathcal{H}_{C}$. Here, we
also define the position and velocity operators, as they will be necessary
to give a physical interpretation to the Poincar\'e algebra. We then
proceed to find an irreducible unitary representation of the Poincar\'e
group that is equivalent to the dynamics of a free particle. The extension
to the case of a particle interacting with an external force is also
given.

In section 4 we give the momentum-based alternative representation
of the Poincar\'e group. We will show that the change to phase space
variables entails a quantum canonical transformation, just as in the
non-relativistic case. The developments of this section leads directly
to the KvN theory for a relativistic (spinless) particle.

Section 5 is devoted to show the close relation of the operational
theory constructed in the previous sections with relativistic Hamiltonian
mechanics.

Finally, in section 6 we investigate the transformation properties,
under Lorentz boosts, of the operators used in the theory. It will
be found that the momentum and velocity operators transform as expected
from relativistic physics. However, the position operator will not
transform as the coordinates do in special relativity, instead it
will transform as a Newton-Wigner operator.

The Einstein summation convention is used throughout this paper. We
work in units where $c=1$.

\section{The proper Poincar\'e group and algebra}

The proper Poincar\'e group is a ten parameter group that consists of
space and time translations, rotations, and Lorentz transformation
(boosts). The generators of these space-time transformations will
be associated with Hermitian operators as follows: $\widehat{\mathcal{J}}_{i}$
will be the generator of rotations around $i$-axis; $\widehat{\lambda}_{x_{i}}$
stands for the space displacement generator in the $i$ -direction;
$\widehat{\mathcal{K}}_{i}$ denotes the Lorentz boost along the $i$-axis;
$\widehat{L}$ stands for the time displacement generator. All of
these operators will act on a Hilbert space to be described in the
next section. The space-time transformations of the Poincar\'e group,
not including the Lorentz boosts, will be realized by unitary operators
with the following convention:

\begin{eqnarray*}
\mathrm{\mathbf{Space-Time\:Transformations}} &  & \mathbf{Unitary\:Operator}\\
\mathrm{Rotation\,around\,the\,\mathit{i}\,axis} & \qquad & \mathrm{}\\
\mathbf{x}\rightarrow R_{i}(\theta)\mathbf{x} & \qquad & e^{-i\theta\widehat{\mathcal{J}}_{i}}.\\
\mathrm{Spatial\:displacement}\\
\mathbf{x}\rightarrow\mathbf{x}+\mathbf{a} &  & e^{-ia_{i}\widehat{\lambda}_{x_{i}}}.\\
\mathrm{Time\:displacement}\\
t\longrightarrow t+\tau &  & e^{i\tau\widehat{L}}.
\end{eqnarray*}

The operator that generates a finite Lorentz transformation to a moving
frame with speed $v$ along the $i$-axis will be denoted by

\begin{equation}
e^{is\widehat{\mathcal{K}}_{i}},
\end{equation}
where $s=\tanh^{-1}v$ is the rapidity.

The above operators are postulated to obey the commutation relations
from the Lie algebra of the proper Poincar\'e group. The derivation
of the Poincar\'e algebra can be found in, for example, \cite{weinberg}.
The commutator relations not involving the time evolution operator
are

\begin{subequations}

\begin{eqnarray}
\left[\widehat{\lambda}_{x_{i}},\widehat{\lambda}_{x_{j}}\right] & = & 0,\label{lamdax}\\
\left[\widehat{\mathcal{J}}_{i},\widehat{\mathcal{J}}_{j}\right] & = & i\varepsilon_{ijk}\widehat{\mathcal{J}}_{k},\label{jj}\\
\left[\widehat{\mathcal{J}}_{i},\widehat{\lambda}_{x_{j}}\right] & = & i\varepsilon_{ijk}\widehat{\lambda}_{x_{k}},\label{jlambdax}\\
\left[\widehat{\mathcal{J}}_{i},\widehat{\mathcal{K}}_{j}\right] & = & i\varepsilon_{ijk}\widehat{\mathcal{K}}_{k},\\
\left[\widehat{\mathcal{K}}_{i},\widehat{\mathcal{K}}_{j}\right] & = & -i\varepsilon_{ijk}\widehat{\mathcal{J}}_{k}.\label{kj}
\end{eqnarray}

\end{subequations}

The commutators that do involve $\widehat{L}$ are

\begin{subequations}

\begin{eqnarray}
\left[\widehat{\mathcal{J}}_{\alpha},\widehat{L}\right] & = & 0,\label{JL}\\
\left[\widehat{\mathcal{K}}_{i},\widehat{L}\right] & = & i\widehat{\lambda}_{x_{i}},\label{gl}\\
\left[\widehat{\lambda}_{x_{i}},\widehat{L}\right] & = & 0,\label{lambdaxL}\\
\left[\widehat{\mathcal{K}}_{i},\widehat{\lambda}_{x_{j}}\right] & = & i\delta_{ij}\widehat{L}.\label{KL}
\end{eqnarray}

\end{subequations}

In the operational version of classical dynamics (relativistic or
not), the generator of the space-time transformations do not have
the same physical interpretation as in quantum mechanics. For example,
the generator of rotation $\widehat{\mathcal{J}}$ is not an angular
momentum operator, and $\widehat{L}$ is not an energy operator. For
this reason, we are using the symbols $(\widehat{\mathcal{J}},\,\widehat{\lambda}_{\mathbf{r}},\,\widehat{\mathcal{K}},\,\widehat{L},)$
instead of the more familiar $(\widehat{\mathbf{J}},\,\widehat{\mathbf{P}},\,\widehat{\mathbf{K}},\,\widehat{H})$.

We will see in section 5 the relation between the time evolution operator
$\widehat{L}$ and the Liouville equation of classical statistical
mechanics; for this reason, we will call $\widehat{L}$ as the Liouvillian
operator.

\section{Classical representation of the Poincar\'e Algebra}

In classical relativistic dynamics there is no uncertainty principle
between the position $\mathbf{r}$ and the velocity $\mathbf{v}$
of a particle, both quantities are postulated to be measurable as
precisely as desired. This lack of uncertainty principle will be formulated
with a commutation relation in Eq. (\ref{XV}). The state of the classical
particle will be described by a vector in a suitable Hilbert space
$\mathcal{H}_{C}$, and these states have to contain information of
the position and velocity of the particle. The situation is almost
the same as in the non-relativistic case, with the only exception
that, here, the speed should not be allowed to be greater than $1$
for physically acceptable states. The vectors $\left|\psi\right\rangle \in\mathcal{H}_{C}$
are postulated to be of the form 

\begin{equation}
\left|\psi\right\rangle =\int\left\langle \mathbf{r},\mathbf{v}\right|\left.\psi\right\rangle \left|\mathbf{r},\mathbf{v}\right\rangle \,d\mathbf{r}d\mathbf{v},\label{ket}
\end{equation}
such that $\psi(\mathbf{r},\mathbf{v})=\left\langle \mathbf{r},\mathbf{v}\right.\left|\psi\right\rangle $
is a square integrable function, and the kets $\left|\mathbf{r},\mathbf{v}\right\rangle $
obey the orthornormality condition

\begin{equation}
\left\langle \mathbf{r}',\mathbf{v}'\right.\left|\mathbf{r},\mathbf{v}\right\rangle =\delta(\mathbf{r}-\mathbf{r}')\delta(\mathbf{v}-\mathbf{v}').
\end{equation}
For physically acceptable states, the limits of integration in (\ref{ket})
are $\left(-\infty,\infty\right)$ for each coordinate and $\left(-1,1\right)$
for each component of the velocity. However, the formalism to be developed
below requires that we accept in $\mathcal{H}_{C}$ states with values
of the speed greater than 1. Nevertheless, we will see that the unitary
time evolution is such that if an state is physically acceptable at
$t=0$, it remains that way for all times. 

The theory will be statistical in nature (though it can describe particle
trajectories if non-normalizable states are allowed, see section 6).
The probability $P(\mathbf{r},\mathbf{v})$ of finding the particle
with position $\mathbf{r}$ and velocity $\mathbf{v}$ is given by
the Born rule

\begin{equation}
P(\mathbf{r},\mathbf{v})=\left|\left\langle \mathbf{r},\mathbf{v}\right.\left|\Psi\right\rangle \right|^{2}.
\end{equation}

The position $\widehat{\mathbf{R}}=\left(\widehat{X}_{1},\widehat{X}_{2},\widehat{X}_{3}\right)$
and velocity $\widehat{\mathbf{V}}=\left(\widehat{V}_{1},\widehat{V}_{2},\widehat{V}_{3}\right)$
operators are defined by their action on the base kets

\begin{eqnarray}
\widehat{X}_{i}\left|\mathbf{r},\mathbf{v}\right\rangle  & = & x_{i}\left|\mathbf{r},\mathbf{v}\right\rangle ,\\
\widehat{V}_{j}\left|\mathbf{r},\mathbf{v}\right\rangle  & = & v_{j}\left|\mathbf{r},\mathbf{v}\right\rangle .
\end{eqnarray}

In order to form an irreducible set of operators in the Hilbert space
we are considering, we define the velocity-translation operator $\widehat{\lambda}_{\mathbf{v}}=\left(\widehat{\lambda}_{v_{1}},\widehat{\lambda}_{v_{2}},\widehat{\lambda}_{v_{3}}\right)$
by

\begin{equation}
e^{-i\mathbf{b}\cdot\widehat{\lambda}_{\mathbf{v}}}\left|\mathbf{r},\mathbf{v}\right\rangle =\left|\mathbf{r},\mathbf{v}+\mathbf{b}\right\rangle .\label{DEf lambdav}
\end{equation}
The translation produced by $\widehat{\lambda}_{\mathbf{v}}$ can
increase the speed above 1, but we will find that this have no negative
effect on the rest of the theory. 

The translation operators $\widehat{\lambda}_{\mathbf{r}}$ and $\widehat{\lambda}_{\mathbf{v}}$
are conjugated in the quantum sense to $\widehat{\mathbf{R}}$ and
$\widehat{\mathbf{V}}$, respectively\footnote{The interpretation of $\widehat{\lambda}_{\mathbf{r}}$ and $\widehat{\lambda}_{\mathbf{v}}$
is not entirely clear. Usually, they are understood to be non-obserbables
or hidden variables \cite{sudarshan-2}. However, it is tempting to
consider them as observables because (1) they are related to ergodic
properties \cite{arnold}, (2) to avoid superselection rules \cite{Cla 8},
and (3) they are useful to construct a classical measurement theory
\cite{cla11}. Moreover, these operator are related to the Bopp operators in the Wigner phase space representation of quantum mechanics \cite{bopp}. }. The following commutation relations are postulated to be satisfied

\begin{subequations}

\begin{eqnarray}
\left[\widehat{X}_{\alpha},\widehat{X}_{\beta}\right] & = & \left[\widehat{X}_{\alpha},\widehat{V}_{\beta}\right]=\left[\widehat{V}_{\alpha},\widehat{V}_{\beta}\right]=0,\label{XV}\\
\left[\widehat{X}_{\alpha},\widehat{\lambda}_{v_{\beta}}\right] & = & \left[\widehat{V}_{\alpha},\widehat{\lambda}_{x_{\beta}}\right]=\left[\widehat{\lambda}_{x_{\alpha}}\widehat{\lambda}_{v_{\alpha}}\right]=0,\\
\left[\widehat{X}_{\alpha},\widehat{\lambda}_{x_{\beta}}\right] & = & \left[\widehat{V}_{\alpha},\widehat{\lambda}_{v_{\beta}}\right]=i\delta_{\alpha\beta}.\label{delta}
\end{eqnarray}

\end{subequations}

Furthermore, $\widehat{\mathbf{R}}$, $\widehat{\mathbf{V}}$, and
$\widehat{\lambda}_{\mathbf{v}}$ are postulated to rotate as vector
operators

\begin{eqnarray}
\left[\widehat{\mathcal{J}}_{i},\widehat{X}_{j}\right] & = & i\varepsilon_{ijk}\widehat{X}_{k},\label{jx}\\
\left[\widehat{\mathcal{J}}_{i},\widehat{V}_{j}\right] & = & i\varepsilon_{ijk}\widehat{V}_{k},\label{jv}\\
\left[\widehat{\mathcal{J}}_{i},\widehat{\lambda}_{v_{j}}\right] & = & i\varepsilon_{ijk}\widehat{\lambda}_{v_{k}}.\label{jlambdav}
\end{eqnarray}

The effect of rotation and space displacement on the base kets will
be\footnote{We will not postulate the effect of Lorentz transformation on $\widehat{\mathbf{R}}$
or on $\left|\mathbf{r},\mathbf{v}\right\rangle $. The transformation
equation will be found a posteriori once the form of the generators
is known.}

\begin{eqnarray}
e^{-i\mathbf{a}\cdot\widehat{\lambda}_{\mathbf{r}}}\left|\mathbf{r},\mathbf{v}\right\rangle  & = & \left|\mathbf{r}+\mathbf{a},\mathbf{v}\right\rangle ,\label{Def lambdar}\\
e^{-i\theta\hat{\mathbf{n}}\cdot\widehat{\mathbf{\mathcal{J}}}}\left|\mathbf{r},\mathbf{v}\right\rangle  & = & \left|\mathbf{r}+\theta\hat{\mathbf{n}}\times\mathbf{r},\mathbf{v}+\theta\hat{\mathbf{n}}\times\mathbf{v}\right\rangle .\label{Def J}
\end{eqnarray}
The effect of $\widehat{L}$ is temporal displacement in the state
vectors, namely

\begin{equation}
e^{-it\widehat{L}}\left|\Psi(0)\right\rangle =\left|\Psi(t)\right\rangle .\label{def time evolution}
\end{equation}
Equation (\ref{def time evolution}) implies a Schr\"{o}dinger-like equation

\begin{equation}
\frac{d}{dt}\left|\Psi(t)\right\rangle =-i\widehat{L}\left|\Psi(t)\right\rangle .\label{defL2}
\end{equation}

As in the non-relativistic quantum \cite{ballentine} and classical
cases \cite{galilei}, $\widehat{\mathbf{R}}$ and $\widehat{\mathbf{V}}$
will be related by

\begin{equation}
\frac{d}{dt}\left\langle \widehat{\mathbf{R}}\right\rangle =\left\langle \widehat{\mathbf{V}}\right\rangle ,\label{RV1}
\end{equation}
or, equivalently, by

\begin{equation}
\widehat{\mathbf{V}}=i\left[\widehat{L},\widehat{\mathbf{R}}\right].\label{RV2}
\end{equation}

Unfortunately, under the unitary transformation generated by $\widehat{\mathcal{K}}$,
the position operator $\widehat{\mathbf{R}}$ cannot be made to transform
as expected from a Lorentz transformation (see section 6). However,
$\widehat{\mathbf{V}}$ can be made to obey the velocity addition
formula. For example, a boost with rapidity $s=\tanh^{-1}v$   in the $z$-axis will give

\begin{eqnarray}
e^{is\widehat{\mathcal{K}}_{z}}\widehat{V}_{x}e^{-is\widehat{\mathcal{K}}_{z}} & = & \frac{\left(1-v^{2}\right)^{1/2}\widehat{V}_{x}}{1-v\widehat{V}_{z}},\label{k1}\\
e^{is\widehat{\mathcal{K}}_{z}}\widehat{V}_{y}e^{-is\widehat{\mathcal{K}}_{z}} & = & \frac{\left(1-v^{2}\right)^{1/2}\widehat{V}_{y}}{1-v\widehat{V}_{z}},\label{k2}\\
e^{is\widehat{\mathcal{K}}_{z}}\widehat{V}_{z}e^{-is\widehat{\mathcal{K}}_{z}} & = & \frac{\widehat{V}_{z}-v}{1-v\widehat{V}_{z}}.\label{k3}
\end{eqnarray}

To end this section we point out that  the operators $\left\{ \widehat{\mathbf{R}},\widehat{\mathbf{V}}\right\} $ form a complete set of commuting observables in $\mathcal{H}_{C}$. This is,  any operator $\widehat{A}$ that commutes with all the elements of $\left\{ \widehat{\mathbf{R}},\widehat{\mathbf{V}}\right\} $  is of the form $\widehat{A}=\widehat{A}(\widehat{\mathbf{R}},\widehat{\mathbf{V}})$.  Physically, this represents the fact that the particle we are considering has no internal degrees of freedom. Now, due to the commutation relations
(\ref{XV}) to (\ref{delta}), no operator $\widehat{A}(\widehat{\mathbf{R}},\widehat{\mathbf{V}})$ can commute with all the elements of $\left\{\widehat{\lambda}_{\mathbf{r}},\widehat{\lambda}_{\mathbf{v}}\right\} $. Hence, only a multiple of the identity can commute with all the elements of the set $\left\{ \widehat{\mathbf{R}},\widehat{\mathbf{V}},\widehat{\lambda}_{\mathbf{r}},\widehat{\lambda}_{\mathbf{v}}\right\}$. Let us remind here the Schur's lemma \cite{jordan}: \emph{A set of self-adjoint operators is irreducible if and only if any operator that commutes with all members of the set is a multiple of the identity}. Hence, we conclude that $\left\{ \widehat{\mathbf{R}},\widehat{\mathbf{V}},\widehat{\lambda}_{\mathbf{r}},\widehat{\lambda}_{\mathbf{v}}\right\}$  is an irreducible set in $\mathcal{H}_{C}$. In the next subsection we will give a realization of the Poincar\'e algebra in terms of $\left\{ \widehat{\mathbf{R}},\widehat{\mathbf{V}},\widehat{\lambda}_{\mathbf{r}},\widehat{\lambda}_{\mathbf{v}}\right\}$, this means that we will have, via the exponential map, an irreducible unitary representation of the Poincar\'e group.

\subsection{Free particle}

We can easily find a suitable realization for $\widehat{\mathcal{J}}$
and $\widehat{L}$ . The operator 

\begin{equation}
\widehat{\mathcal{J}}_{k}=\varepsilon_{ijk}\left(\widehat{X}_{j}\widehat{\lambda}_{x_{k}}+\widehat{V}_{j}\widehat{\lambda}_{v_{k}}\right),\label{formJ}
\end{equation}
satisfy Eqs. (\ref{jj}), (\ref{jlambdax}), (\ref{jx}), (\ref{jv}),
(\ref{jlambdav}) and (\ref{Def J}). On the other hand, we expect
the wave functions of free particles to evolve in time according to
$\psi(\mathbf{r},\mathbf{v})\rightarrow\psi(\mathbf{r}-\mathbf{v}t,\mathbf{v})$,
and that is accomplished if the effect of the Liouvillian on the base
kets is given by

\begin{equation}
e^{-it\widehat{L}}\left|\mathbf{r},\mathbf{v}\right\rangle =\left|\mathbf{r}+\mathbf{v}t,\mathbf{v}\right\rangle .\label{free}
\end{equation}
A viable solution to Eq. (\ref{free}) is

\begin{equation}
\widehat{L}=\widehat{V}_{i}\widehat{\lambda}_{x_{i}}.\label{Lfree}
\end{equation}
The operator (\ref{Lfree}) is invariant under translations and rotations,
as it should be from the Poincar\'e algebra. $\widehat{\mathcal{J}}$
and $\widehat{L}$ have the same form as the generator of the Galilei
algebra for the non-relativistic case. 

The only non-trivial task is to find the generators of Lorentz boosts
$\widehat{\mathcal{K}}$, however, trial and error give the suitable
operator

\begin{align}
\widehat{\mathcal{K}}_{i} & =\left(\widehat{X}_{i}\widehat{L}-\left\{ \delta_{ij}-\widehat{V}_{i}\widehat{V}_{j}\right\} \widehat{\lambda}_{v_{j}}\right)_{S}-t\widehat{\lambda}_{x_{i}},\label{K}
\end{align}
where we used the notation $\left(\widehat{A}\widehat{B}\right)_{S}=\frac{1}{2}\left(\widehat{A}\widehat{B}+\widehat{B}\widehat{A}\right)$.
There is, actually, some freedom in the choice of $\widehat{\mathcal{K}}$.
We can remove the term $t\widehat{\lambda}_{x_{i}}$ without affecting
the commutation relations of the Poincar\'e algebra. However, it is
preferable to keep this term as it enhances the physical interpretation
of the generator of the Lorentz transformations, as we will see in
section 5.

Under the Schr\"{o}dinger evolution
\begin{equation}
i\frac{\partial}{\partial t}\psi(\mathbf{r},\mathbf{v},t)=\widehat{L}\psi(\mathbf{r},\mathbf{v},t),\label{sc}
\end{equation}
 a wave function evolve as expected from a free particle $\psi(\mathbf{r},\mathbf{v})\rightarrow\psi(\mathbf{r}-\mathbf{v}t,\mathbf{v})$.
Moreover, the Heisenberg evolution of $\widehat{\mathbf{R}}$ and
$\widehat{\mathbf{V}}$ matches the behavior of a free particle 

\begin{align}
\frac{d}{dt}\widehat{\mathbf{R}}(t) & =i\left[\widehat{L},\widehat{\mathbf{R}}\right]=\widehat{\mathbf{V}},\nonumber \\
\frac{d}{dt}\widehat{\mathbf{V}}(t) & =i\left[\widehat{L},\widehat{\mathbf{V}}\right]=0.
\end{align}

To summarize, In view of the relations (\ref{XV}) to (\ref{delta}),
the operators $\widehat{\mathcal{K}}$, $\widehat{\mathcal{J}}$ and
$\widehat{L}$ given by (\ref{formJ}), (\ref{Lfree}), and (\ref{K}),
respectively, satisfy all the commutation relations of the Poincar\'e
algebra. They generate a representation of the Poincar\'e group that
is irreducible in the Hilbert space we are considering, and this representation
gives the correct dynamics of a free particle.

\subsection{Interaction with an external field}

The generators $\widehat{\mathcal{J}}$ and $\widehat{\mathcal{K}}$
are understood to be geometrical in nature, they do not change in
the presence of interactions. Hence, the relations (\ref{lamdax})
to (\ref{kj}) remain valid.

$\widehat{L}$, on the other hand, is now to be understood as the
generator of a \emph{dynamical }evolution in time, and its form change
accordingly. Thus, relations (\ref{JL}) to (\ref{KL}) could get
modified. The equation of motion is still postulated to be

\begin{eqnarray}
\frac{d}{dt}\left|\Psi(t)\right\rangle  & = & -i\widehat{L}\left|\Psi(t)\right\rangle .\label{defL2-1}
\end{eqnarray}

In the non-relativistic case, $\widehat{L}$ is identified by demanding
that the acceleration is independent of the unobservable operators
$\widehat{\lambda}_{\mathbf{r}}$ and $\widehat{\lambda}_{\mathbf{v}}$,
this is, the acceleration obeys the Newton equation 
\begin{equation}
\frac{d}{dt}\widehat{\mathbf{V}}(t)=\frac{1}{m}\widehat{\mathbf{F}}(\widehat{\mathbf{R}},\widehat{\mathbf{V}}),
\end{equation}
for some $\widehat{\mathbf{F}}$. The mass $m$ and the force operator
$\widehat{\mathbf{F}}$ are to be found from experiments. 

We can make a similar demand to find $\widehat{L}$ in the relativistic
case. It can be checked that 

\begin{equation}
\widehat{L}=\widehat{V}_{i}\widehat{\lambda}_{x_{i}}+\left(\frac{1}{m_{0}}\widehat{\gamma}^{-1}\left\{ F_{i}-(\widehat{\mathbf{V}}\cdot\widehat{\mathbf{F}})\widehat{V}_{i}\right\} \widehat{\lambda}_{v_{i}}\right)_{S},\label{Lnew}
\end{equation}
where $\widehat{\gamma}=\left(1-\widehat{\mathbf{V}}^{2}\right)^{-1}$,
reproduces the relativistic force equation
\begin{equation}
m_{0}\frac{d}{dt}\left(\widehat{\gamma}\widehat{\mathbf{V}}(t)\right)=\widehat{\mathbf{F}}(\widehat{\mathbf{R}},\widehat{\mathbf{V}}).\label{Rforce}
\end{equation}

\subsubsection*{Example: Constant force.}

We can take the case of a constant force as an example. For simplicity,
we can consider the one-dimensional case of a particle restricted
to move in the $x$-axis. In this case, the Liouvillian reduces to

\begin{equation}
\widehat{L}=\widehat{V}\widehat{\lambda}_{x}+\frac{F}{m_{0}}\left(\left\{ 1-\widehat{V}^{2}\right\} ^{3/2}\widehat{\lambda}_{v}\right)_{S}.\label{Lnew-1}
\end{equation}
The Heisenberg equation of motion for $\widehat{V}$ can be computed
to give

\begin{equation}
\frac{d}{dt}\widehat{V}=i\left[\widehat{L},\widehat{V}\right]=\frac{F}{m_{0}}\left\{ 1-\widehat{V}^{2}\right\} ^{3/2}.\label{dV}
\end{equation}
Equation (\ref{dV}) can be inverted and integrated to obtain $\widehat{V}(t)$
and $\widehat{X}(t)$. Assuming $\widehat{V}(0)=0$ and $\widehat{X}(0)=0$,
we have

\[
\widehat{V}(t)=\frac{\frac{1}{m_{0}}Ft}{\sqrt{1+\left(\frac{Ft}{m_{0}}\right)^{2}}},
\]

and

\[
\widehat{X}(t)=\frac{F}{m_{0}}\left(\sqrt{1+\left(\frac{Ft}{m_{0}}\right)^{2}}-1\right).
\]

We can see that the operators $\widehat{V}(t)$ and $\widehat{X}(t)$
are as expected compared to the known behavior of this system in relativistic
dynamics. Moreover, knowing the solution to the Heisenberg equation,
we can pass to the Scrh\"{o}dinger picture and write the evolution of
basis kets as

\begin{equation}
e^{-it\widehat{L}}\left|x,v\right\rangle =\left|x(t),v(t)\right\rangle ,
\end{equation}
where $x(t)=\frac{F}{m_{0}}\sqrt{1+\left(\frac{Ft}{m_{0}}\right)^{2}}-\frac{F}{m_{0}}$
and $v(t)=\frac{Ft}{m_{0}}\left[1+\left(\frac{Ft}{m_{0}}\right)^{2}\right]^{-1/2}.$
For basis kets with $v<1$, the time evolution guarantees that $v(t)<1$
for all times. Hence, a physically acceptable state will remain so
for all times. This result has been obtained for the particular case
of a constant force in one dimension, however, it is a general result
since, by construction, the Heisenberg evolution of the operators
$\widehat{\mathbf{R}}(t)$ and $\widehat{\mathbf{V}}(t)$ mimic the
behavior of $\mathbf{r}(t)$ and $\mathbf{v}(t)$ obtained from the
relativistic force equation.

\section{Lagrangian Operator}

It is posible to write the elements of the Poincar\'e algebra in terms
of a canonical momentum operator instead of the velocity, obtaining,
then, the relativistic KvN theory. 
To define the momentum operator we  first have to write the equation of motion in the Euler-Lagrange
form. Equation (\ref{Rforce}) can be rewritten as 

\begin{equation}
-\left[\widehat{L},\left[\widehat{\lambda}_{v_{i}},\widehat{T}^{*}\right]\right]=\widehat{F}_{i},\label{LVF-3}
\end{equation}
where the kinetic co-energy \cite{co} is given by $\widehat{T}^{*}=m_{0}[1-\left(1-\widehat{V}^{2}\right)^{1/2}]$.
So far, we were not worried about the nature of the $\widehat{\mathbf{F}}$.
Relativistic principles restrict the possible 3-forces depending on
the tensor decomposition of the 4-force \cite{barut}. We will focus
our attention to Lorentz-like forces. This kind of forces can be obtained
from a generalized potential \cite{galilei}
\begin{align}
\widehat{U}\left(\widehat{\mathbf{R}},\widehat{\mathbf{V}}\right) & =\widehat{\phi}(\widehat{\mathbf{R}})-\widehat{V}_{i}\widehat{A}_{i}(\widehat{\mathbf{R}}),\label{Gpotential}
\end{align}
according to

\begin{equation}
\widehat{F}_{k}=-i\left[\widehat{\lambda}_{x_{k}},\widehat{U}\right]-\left[\widehat{L},\left[\widehat{\lambda}_{v_{k}},\widehat{U}\right]\right]=\widehat{E}_{k}+\left(\widehat{\mathbf{V}}\times\widehat{\mathbf{B}}\right)_{k},\label{fpotential}
\end{equation}
where

\begin{eqnarray}
\widehat{E}_{k} & = & -\frac{\partial\widehat{\phi}}{\partial\widehat{X}_{k}}-\frac{\partial\widehat{A}_{k}}{\partial t},\\
\widehat{B}_{k} & = & \left(\nabla\times\widehat{\mathbf{A}}\right)_{k}.
\end{eqnarray}

The relativistic force equation can then be rewritten as 
\begin{equation}
\Phi[\widehat{\mathcal{L}}]=0,\label{Euler-Lagrange}
\end{equation}
where the Lagrangian operator is

\begin{equation}
\widehat{\mathcal{L}}=\widehat{T}^{*}-\widehat{U},\label{LAGRANGIAN}
\end{equation}
and $\Phi$ is the superoperator defined by

\begin{equation}
\Phi=-\left[\widehat{L},\left[\widehat{\lambda}_{v_{\alpha}},\right]\right]-i\left[\widehat{\lambda}_{x_{\alpha}},\right].\label{superoperator}
\end{equation}

\subsection{Momentum Representation}

As in the nonrelativistic case \cite{galilei}, the Lagrangian operator
is used to define the canonical momentum operator by $\widehat{\mathbf{P}}=i\left[\widehat{\lambda}_{\mathbf{v}},\widehat{\mathcal{L}}\right]$.
Using Eq. (\ref{LAGRANGIAN}), we obtain

\begin{equation}
\widehat{\mathbf{P}}=m_{0}\widehat{\gamma}\widehat{\mathbf{V}}+\widehat{\mathbf{A}}.\label{defP}
\end{equation}
The following commutation relations for $\widehat{\mathbf{P}}$ are
satisfied

\begin{eqnarray}
\left[\widehat{X}_{i},\widehat{P}_{j}\right] & = & 0,\\
\left[\widehat{P}_{i},\widehat{\lambda}_{x_{j}}\right] & = & i\frac{\partial\widehat{A}_{i}}{\partial\widehat{X}_{j}},\label{a1}\\
\left[\widehat{P}_{i},\widehat{\lambda}_{v_{j}}\right] & = & im_{0}\widehat{\gamma}\left(\delta_{ij}+V_{i}V_{j}\widehat{\gamma}\right).\label{a2}
\end{eqnarray}

We can now make the definitions
\begin{eqnarray}
\widehat{\lambda}_{p_{i}} & = & \frac{1}{m_{0}}\left[\widehat{\gamma}^{-1}\left(\widehat{\lambda}_{v_{i}}-V_{i}V_{k}\widehat{\lambda}_{v_{k}}\right)\right]_{S},\label{changelambda1}\\
\widehat{\lambda}'_{x_{j}} & = & \widehat{\lambda}_{x_{j}}-\frac{\partial\widehat{A}_{i}}{\partial\widehat{X}_{j}}\widehat{\lambda}_{p_{i}},\label{changelambda2}
\end{eqnarray}
and, then, write the commutation relations for the set $\left\{ \widehat{\mathbf{R}},\widehat{\mathbf{P}},\widehat{\lambda}'_{\mathbf{r}},\widehat{\lambda}_{\mathbf{p}}\right\} $
as

\begin{subequations}

\begin{eqnarray}
\left[\widehat{X}_{i},\widehat{X}_{_{j}}\right] & = & \left[\widehat{X}_{i},\widehat{P}_{j}\right]=\left[\widehat{P}_{i},\widehat{P}_{j}\right]=0,\label{Xp}\\
\left[\widehat{X}_{i},\widehat{\lambda}_{p_{i}}\right] & = & \left[\widehat{P}_{\alpha},\widehat{\lambda}'_{x_{\beta}}\right]=\left[\widehat{\lambda}'_{x_{j}},\widehat{\lambda}_{p_{i}}\right]=0,\\
\left[\widehat{X}_{i},\widehat{\lambda}'_{x_{j}}\right] & = & \left[\widehat{P}_{i},\widehat{\lambda}_{p_{j}}\right]=i\delta_{ij}.\label{Xlambda-1}
\end{eqnarray}

\end{subequations}

The set of operators $\left\{ \widehat{\mathbf{R}},\widehat{\mathbf{P}},\widehat{\lambda}'_{\mathbf{r}},\widehat{\lambda}_{\mathbf{p}}\right\} $
is irreducible in the Hilbert space we are considering in view of
the preceding set of equations (\ref{Xp}) to (\ref{Xlambda-1}).
We can write the elements of the Poincar\'e algebra $(\widehat{\mathcal{J}},\,\widehat{\mathcal{K}},\,\widehat{L})$
in terms of $\left\{ \widehat{\mathbf{R}},\widehat{\mathbf{P}},\widehat{\lambda}'_{\mathbf{r}},\widehat{\lambda}_{\mathbf{p}}\right\} $,
obtaining, then, a momentum-based irreducible representation of the
Poincar\'e group. This alternative representation is gauge dependent
due to the presence of the vector potential $\widehat{\mathbf{A}}$.
We can give an example for the free particle, defining 
\begin{equation}
\mathcal{\widehat{H}}=\sqrt{\widehat{\mathbf{P}}^{2}+m_{0}^{2}},
\end{equation}
the elements of the algebra becomes

\begin{align}
\widehat{\mathcal{J}}_{k} & =\varepsilon_{ijk}\left(\widehat{X}_{j}\widehat{\lambda}'_{x_{k}}+\widehat{P}_{j}\widehat{\lambda}_{p_{k}}\right),\label{JP}\\
\widehat{L}' & =\widehat{P}_{i}\mathcal{\widehat{H}}^{-1}\widehat{\lambda}'_{x_{i}},\label{LP}\\
\widehat{\mathcal{K}}_{i} & =\left(\widehat{X}_{i}\widehat{L}-\mathcal{\widehat{H}}\,\widehat{\lambda}_{p_{i}}\right)_{S}-t\widehat{\lambda}'_{x_{i}}.\label{KP}
\end{align}

In presence of a general Lorentz interactions, the general form of the Liouvillian is
given by the rather complicated expresion
\begin{align}
\widehat{L}' & =\left[\mathcal{\widehat{H}}^{-1}\left(\widehat{P}_{i}-\widehat{A}_{i}\right)\left(\widehat{\lambda}'_{x_{i}}+\frac{\partial\widehat{A}_{i}}{\partial\widehat{X}_{j}}\widehat{\lambda}_{p_{i}}\right)+\widehat{\mathbf{F}}\cdot\widehat{\lambda}_{\mathbf{p}}\right.\nonumber \\
 & -\mathcal{\widehat{H}}^{-2}\left\{ \widehat{A}_{i}\left(\widehat{\mathbf{F}}\cdot\widehat{\mathbf{P}}\right)+\left(\widehat{P}_{i}-\widehat{A}_{i}\right)\left(\widehat{\mathbf{F}}\cdot\widehat{\mathbf{A}}\right)\right\} \nonumber \\
 & .\left.\times\left(\delta_{ij}-\widehat{P}_{i}\widehat{A}_{j}-\widehat{P}_{j}\widehat{A}_{i}+\widehat{A}_{i}\widehat{A}_{j}\right)\widehat{\lambda}_{p_{j}}\right]_{S}.\label{LA}
\end{align}
where $\widehat{\mathbf{F}}$ is the Lorentz force operator given
in (\ref{fpotential}) and now $\mathcal{\widehat{H}}=m_{0}\widehat{\gamma}=\sqrt{\left(\widehat{\mathbf{P}}-\widehat{\mathbf{A}}\right)^{2}+m_{0}^{2}}$.
Equation (\ref{LA}) is the relativistic generalization of the minimal-coupling
rule for the KvN theory given in \cite{minimal coupling}. For purely
electric fields ($\widehat{\mathbf{A}}=0$), the Liouvillian (\ref{LA})
considerably simplifies to

\begin{equation}
\widehat{L}'=\mathcal{\widehat{H}}^{-1}\widehat{\mathbf{P}}\cdot\widehat{\lambda}'_{\mathbf{r}}+\widehat{\mathbf{F}}\cdot\widehat{\lambda}_{\mathbf{p}}.\label{LA2}
\end{equation}

The change from $\left\{ \widehat{\mathbf{R}},\widehat{\mathbf{V}},\widehat{\lambda}_{\mathbf{r}},\widehat{\lambda}_{\mathbf{v}}\right\} $
to $\left\{ \widehat{\mathbf{R}},\widehat{\mathbf{P}},\widehat{\lambda}'_{\mathbf{r}},\widehat{\lambda}_{\mathbf{p}}\right\} $
is a quantum canonical transformation, i.e., a transformation that
preserve the commutation relations. Unitary transformations are canonical,
but the converse is not necessarily true \cite{QCT}. In the nonrelativistic
case, the pass from velocity representation to momentum representation
is performed by a composition of a scale and a unitary transformation
\cite{galilei}, we will now show that the same is true in the relativistic
case. The transformation $\left\{ \widehat{\mathbf{R}},\widehat{\mathbf{V}},\widehat{\lambda}_{\mathbf{r}},\widehat{\lambda}_{\mathbf{v}}\right\} \rightarrow\left\{ \widehat{\mathbf{R}},\widehat{\mathbf{P}},\widehat{\lambda}'_{\mathbf{r}},\widehat{\lambda}_{\mathbf{p}}\right\} $
can be given in two steps. First, we make the change

\begin{subequations}

\begin{eqnarray}
\widehat{\mathbf{V}} & \longrightarrow & m_{0}\widehat{\mathbf{V}},\\
\widehat{\lambda}_{\mathbf{v}} & \longrightarrow & \frac{1}{m_{0}}\widehat{\lambda}_{\mathbf{v}}.
\end{eqnarray}

\end{subequations}The second step consists in a similarity transformation
by the unitary operator

\begin{subequations}

\begin{align}
\widehat{C} & =\widehat{C}_{2}\widehat{C}_{1},\label{C}\\
\widehat{C}_{1} & =\exp\left[\frac{i}{2}\left(\widehat{V}^{2}\,\widehat{\mathbf{V}}\cdot\lambda_{\mathbf{v}}\right)_{S}\right],\label{C1}\\
\widehat{C}_{2} & =\exp\left[i\mathbf{A}\cdot\lambda_{\mathbf{p}}\right].\label{C2}
\end{align}
\end{subequations}The proof that $\widehat{C}$  leads to the correct
canonical transformation $\left\{ \widehat{\mathbf{R}},m_{0}\widehat{\mathbf{V}},\widehat{\lambda}_{\mathbf{r}},\frac{1}{m_{0}}\widehat{\lambda}_{\mathbf{v}}\right\} \rightarrow\left\{ \widehat{\mathbf{R}},\widehat{\mathbf{P}},\widehat{\lambda}'_{\mathbf{r}},\widehat{\lambda}_{\mathbf{p}}\right\} $
is given in the appendix. We point out that the role of $\widehat{C}_{1}$
is to pass from the nonrelativistic kinematic momentum to the relativistic
one

\begin{equation}
\widehat{C}_{1}\left(m_{0}\widehat{\mathbf{V}}\right)\widehat{C}_{1}^{-1}=m_{0}\widehat{\gamma}\widehat{\mathbf{V}},
\end{equation}
and the role of $\widehat{C}_{2}$ is to go into the canonical momentum
by inclusion of the vector potential
\begin{equation}
\widehat{C}_{2}\left(m_{0}\widehat{\gamma}\widehat{\mathbf{V}}\right)\widehat{C}_{2}^{-1}=m_{0}\widehat{\gamma}\widehat{\mathbf{V}}+\widehat{\mathbf{A}}.
\end{equation}

There are two nonequivalent ways to define phase space kets. We could
define them by the unitary transformation $\left|\mathbf{r},\mathbf{p}\right\rangle =\widehat{C}\left|\mathbf{r},\mathbf{v}\right\rangle $,
as it was done in \cite{galilei}. However, this would result in $\widehat{\mathbf{P}}$
having the undesirable eigenvalue equation $\widehat{\mathbf{P}}\left|\mathbf{r},\mathbf{p}\right\rangle =m_{0}\mathbf{v}\left|\mathbf{r},\mathbf{p}\right\rangle $.
In the next section we will relate the operational theory we have
just developed with the results of Hamiltonian mechanics, the following simple identification is then preferable
\begin{equation}
\left|\mathbf{r},\mathbf{p}\right\rangle \equiv\left|\mathbf{r},\mathbf{v}\right\rangle .\label{rp}
\end{equation}
On $\left|\mathbf{r},\mathbf{p}\right\rangle$, $\widehat{\mathbf{R}}$ and $\widehat{\mathbf{P}}$
act as multiplicative operators

\begin{subequations}

\begin{align}
\widehat{\mathbf{R}}\left|\mathbf{r},\mathbf{p}\right\rangle  & =\mathbf{r}\left|\mathbf{r},\mathbf{p}\right\rangle ,\\
\widehat{\mathbf{P}}\left|\mathbf{r},\mathbf{p}\right\rangle  & =\mathbf{p}\left|\mathbf{r},\mathbf{p}\right\rangle .
\end{align}
\end{subequations}That $\widehat{\lambda}'_{\mathbf{r}}$ and $\widehat{\lambda}_{\mathbf{p}}$
act as translation operator on $\left|\mathbf{r},\mathbf{p}\right\rangle $
follow from the commutation relations (\ref{Xp}) to (\ref{Xlambda-1}). 

The Hilbert space spanned by the kets $\left|\mathbf{r},\mathbf{p}\right\rangle ,$
the set of operators $\left\{ \widehat{\mathbf{R}},\widehat{\mathbf{P}},\widehat{\lambda}'_{\mathbf{r}},\widehat{\lambda}_{\mathbf{p}}\right\} $,
the Liouvillian (\ref{LA}) or (\ref{LA2}), together with the Born
rule $P(\mathbf{r},\mathbf{p})=\left|\left\langle \mathbf{r},\mathbf{v}\right.\left|\Psi\right\rangle \right|^{2}$
give the relativistic generalization of the KvN formulation of classical
mechanic.

\section{Relation with Hamiltonian mechanics}

In this section, we will link the theory developed above with the
usual relativistic Hamiltonian mechanics. It is enough to show the
derivation of Hamiltonian mechanics from the KvN formalism. Consider
the wave function 

\begin{equation}
\psi(\mathbf{r},\mathbf{p})=\left\langle \mathbf{r},\mathbf{p}\right|\left.\psi\right\rangle .
\end{equation}
The position and momentum operators act as multiplication operators
when acting on $\psi(\mathbf{r},\mathbf{p})$

\begin{eqnarray}
\widehat{X}_{i}\psi(\mathbf{r},\mathbf{p}) & = & x_{i}\psi(\mathbf{r},\mathbf{p}),\\
\widehat{P}_{j}\psi(\mathbf{r},\mathbf{p}) & = & p_{j}\psi(\mathbf{r},\mathbf{p}).
\end{eqnarray}
On the other hand, the operators $\widehat{\lambda}_{\mathbf{p}}$
and $\widehat{\lambda}'_{\mathbf{r}}$ act as derivatives 

\begin{eqnarray}
\widehat{\lambda}'_{\mathbf{r}}\psi(\mathbf{r},\mathbf{p}) & = & -i\nabla_{\mathbf{r}}\psi(\mathbf{r},\mathbf{p}),\\
\widehat{\lambda}_{\mathbf{p}}\psi(\mathbf{r},\mathbf{p}) & = & -i\nabla_{\mathbf{p}}\psi(\mathbf{r},\mathbf{p}).
\end{eqnarray}

With the help of the Poisson bracket $\left\{ a,b\right\} =\frac{\partial a}{\partial x_{i}}\frac{\partial b}{\partial p_{i}}-\frac{\partial b}{\partial x_{i}}\frac{\partial a}{\partial p_{i}}$,
we can write the components of $\widehat{\lambda}'_{\mathbf{r}}$
and $\widehat{\lambda}_{\mathbf{p}}$ as

\begin{eqnarray}
\widehat{\lambda}'_{x_{i}} & = & -i\left\{ \,,p_{i}\right\} ,\\
\widehat{\lambda}_{p_{j}} & = & i\left\{ \,,x_{j}\right\} .
\end{eqnarray}
The elements of the Poincar\'e algebra (\ref{JP}) to (\ref{KP}) can
be written as 

\begin{align}
\widehat{L}' & =-i\left\{ \,,H\right\} ,\label{newL-1}\\
\widehat{\mathcal{J}} & =-i\left\{ \,,\mathbf{r}\times\mathbf{p}\right\} ,\\
\widehat{\mathcal{K}} & =-i\left\{ \,,\mathbf{r}H-\mathbf{p}t\right\} ,
\end{align}
where the Hamiltonian function is $H=\sqrt{\mathbf{p}^{2}+m_{0}^{2}}$.
We can see that the elements of the Poincar\'e algebra are related,
but are not identified with, the energy $H$, the angular momentum
$\mathbf{r}\times\mathbf{p}$, and the center of energy $\mathbf{r}H-\mathbf{p}t$.

For a dynamical evolution, the Liouvillian retains the form (\ref{newL-1}),
but the Hamiltonian function changes to

\begin{equation}
H=\sqrt{\left(\mathbf{p}-\mathbf{A}\right)^{2}+m_{0}^{2}}+\phi(\mathbf{r}).
\end{equation}

The Schr\"{o}dinger-like equation (\ref{defL2}) becomes

\begin{equation}
\frac{\partial\psi}{\partial t}+\left\{ \psi,H\right\} =0,\label{Liouvillepsi}
\end{equation}
or written for the complex conjugate $\psi^{*}$

\begin{equation}
\frac{\partial\psi^{*}}{\partial t}+\left\{ \psi^{*},H\right\} =0.\label{Liouvillepsi2}
\end{equation}

Defining the probability density in phase-space $\rho=\left|\psi\right|^{2}$,
equations (\ref{Liouvillepsi}) and (\ref{Liouvillepsi2}) can be
combined to give

\begin{equation}
\frac{\partial\rho}{\partial t}+\left\{ \rho,H\right\} =0.\label{cliouville}
\end{equation}
Equation (\ref{cliouville}) is the classical Liouville equation.
Hence, the operational formulation of classical mechanics is equivalent
to classical statistical mechanics.

Particle dynamics is recovered when we considerer classical pure states,
this is, when the probability density is allowed to be a point-like
distribution

\begin{equation}
\rho(t)=\delta(\mathbf{r}-\mathbf{r}(t))\delta(\mathbf{p}-\mathbf{p}(t)).\label{kilmontovich:}
\end{equation}

\section{Transformation properties}

We can find the transformation properties of an operator under a
finite Lorentz transformation just by computing its commutator with
$\widehat{\mathcal{K}}$. For example, computing the commutators between
$\widehat{\mathcal{K}}$ and $\widehat{\mathbf{V}}$ gives
\begin{equation}
\left[\widehat{\mathcal{K}}_{j},\widehat{V}_{k}\right]=i\left(\delta_{jk}-\widehat{V}_{j}\widehat{V}_{k}\right).\label{KVcomm}
\end{equation}
The commutation relation (\ref{KVcomm}) is also found in the quantum
case \cite{Vjordan}. Finite Lorentz transformation are obtained by
a similarity transformation 
\begin{equation}
e^{is\widehat{\mathcal{K}}_{j}}\widehat{V}_{k}e^{-is\widehat{\mathcal{K}}_{j}}.\label{similarityV}
\end{equation}
The computation of (\ref{similarityV}) reduces to a computation of
nested commutator via the Baker-Campbell-Hausdorff expansion. The
similarity transformation (\ref{similarityV}) has the same outcome
as in the quantum case since the commutator algebras coincide. The
result is that the velocity operator obeys the velocity addition formula.
For example, a boost in the $z$-axis results in the Eqs. (\ref{k1})
to (\ref{k3}).

The set $\left(\mathcal{\widehat{H}},\widehat{\mathbf{P}}\right)$
transform as the elements of a 4-vector. This follows from the brackets

\begin{align}
\left[\widehat{\mathcal{K}}_{i},\widehat{P}_{j}\right] & =\delta_{ij}\mathcal{\widehat{H}},\\
\left[\widehat{\mathcal{K}}_{k},\mathcal{\widehat{H}}\right] & =i\widehat{P}_{k}.
\end{align}
For example, a boost in the $z$-axis gives

\begin{align}
e^{is\widehat{\mathcal{K}}_{z}}\mathcal{\widehat{H}}e^{-is\widehat{\mathcal{K}}_{z}} & =\mathcal{\widehat{H}}\,\cosh b-\widehat{P}_{z}\,\sinh b,\\
e^{is\widehat{\mathcal{K}}_{z}}\widehat{P}_{z}e^{-is\widehat{\mathcal{K}}_{z}} & =\widehat{P}_{z}\,\cosh b-\mathcal{\widehat{H}}\,\sinh b,\\
e^{is\widehat{\mathcal{K}}_{z}}\widehat{P}_{i}e^{-is\widehat{\mathcal{K}}_{z}} & =\widehat{P}_{i}\quad(i=1,2).
\end{align}
The sets $\left(\widehat{L},\widehat{\lambda}_{\mathbf{v}}\right)$
and $\left(\widehat{L}',\widehat{\lambda}_{\mathbf{p}}\right)$ also
transform as a 4-vector in view of the Poincar\'e algebra relations
(\ref{gl}) and (\ref{KL}).

The position operator $\widehat{\mathbf{R}}$ does not transform as
the spatial components of a 4-vector. It is the exact same situation encountered
in relativistic quantum mechanics for the Newton-Wigner operator \cite{newton-wigner,newton-wigner2}
and for the very same reasons. As in quantum theory, the formalism
developed in this paper does not consider space and time on equal
grounds. An Hermitian operator is assigned to position, whereas time
is considered to be just a parameter. It can be shown that $\widehat{\mathbf{R}}$
transforms, up to a symmetrization, as a Newton-Wigner operator. For
example, a Lorentz transformation in the z-axis gives
\begin{align}
e^{is\widehat{\mathcal{K}}_{z}}\widehat{z}e^{-is\widehat{\mathcal{K}}_{z}} & =\gamma^{-1}\widehat{z}+v\gamma^{-1}\left(\frac{\widehat{z}\widehat{V}_{z}}{1-v\widehat{V}_{z}}\right),\label{z}\\
e^{is\widehat{\mathcal{K}}_{z}}\widehat{X}_{i}e^{-is\widehat{\mathcal{K}}_{z}} & =\widehat{X}_{i}+v\left(\frac{\widehat{z}\widehat{V}_{i}}{1-v\widehat{V}_{z}}\right)\quad(i=1,2).\label{xy}
\end{align}

To prove the above statement, is suficcient to show that the right
and left hand sides of the previous equations agree up to the firsts
lowest orders in a series expansion. The equality of the entire expresions
follows since both sides are a one parameter group of transformation.
The left-hand side of Eq.(\ref{z}) can be expanded using the Baker-Campbell-Hausdorf
formula$e^{X}Ye^{-X}=Y+[X,Y]+\frac{1}{2}[X,[X,Y]]+\ldots$, where

\begin{align}
\left[\widehat{\mathcal{K}}_{z},\widehat{z}\right] & =i\left(\widehat{z}\widehat{V}_{z}\right),\\
\left[\widehat{\mathcal{K}}_{z},\left[\widehat{\mathcal{K}}_{z},\widehat{z}\right]\right] & =\widehat{z}-2\widehat{z}\widehat{V}_{z}.
\end{align}
The right hand side of (\ref{z}) can be expanded using $\gamma^{-1}=1-\frac{1}{2}v^{2}-\frac{1}{8}v^{4}\ldots$
and $\frac{1}{1-v\widehat{V}_{z}}=1+v\widehat{V}_{z}+\ldots$. It
follows that both sides agree up to order $v^{2}$, hence, Eq.(\ref{z})
is an identity. A similar reasoning shows the validity of Eq. (\ref{xy}).

\section{Concluding Remarks}

We obtained an operational formulation
of classical relativistic dynamics from a unitary irreducible representation
of the Poincar\'e group, extending the results given in \cite{galilei}
for the Galilei group and non-relativistic mechanics. Moreover, we
have given, from first principles, a relativistic generalization of
the minimal coupling rule in the KvN theory given in \cite{minimal coupling}.
These results are completely independent from quantum mechanics as
no classical limit was employed anywhere.

As in the nonrelativistic case \cite{galilei}, the theory can
be equivalently formulated in terms of the velocity or in terms of
the canonical momentum. The passing from a theory formulated in the
tangent bundle of configuration space to a theory in phase space is
done by a quantum canonical transformation that is a composition
of a scale and a unitary transformation. However, contrary to the
nonrelativistic case, it was found that it is preferable to not
define the kets $\left|\mathbf{r},\mathbf{p}\right\rangle $ with
the same unitary operator used to pass from $\left\{ \widehat{\mathbf{R}},m_{0}\widehat{\mathbf{V}},\widehat{\lambda}_{\mathbf{r}},\frac{1}{m_{0}}\widehat{\lambda}_{\mathbf{v}}\right\} $
to $\left\{ \widehat{\mathbf{R}},\widehat{\mathbf{P}},\widehat{\lambda}'_{\mathbf{r}},\widehat{\lambda}_{\mathbf{p}}\right\} $.
The ad hoc identification $\left|\mathbf{r},\mathbf{p}\right\rangle $
given in Eq. (\ref{rp}) was chosen in order for us to get the relativistic
KvN formalism and, later, relativistic Hamiltonian mechanics.

The majority of the operators we used in this paper transform as expected
from relativistic physics. The sole exception is the position operator
as it transforms not as the spatial components of a 4-vector but as
a Newton-Wigner operator. The relation between our position operator
and the Newton-Wigner function in Hamiltonian mechanics \cite{clanewton}
remains to be explored.

The classification of all the \emph{classical} unitary irreducible
representation of the proper Poincar\'e group remains unsolved as we
have only given one possible realization. In this regard, the physically most relevant open problem is to  find representations that leads to the dynamics  of massless particles and the inclusion of the (classical)
intrinsic angular momentum.

\appendix
\section{Appendix}
\renewcommand{\theequation}{A-\arabic{equation}}
\setcounter{equation}{0} 

The purpose of this appendix is to prove that the quantum canonical
transformation

\begin{align}
\widehat{\mathbf{R}} & \rightarrow\widehat{\mathbf{R}},\\
m_{0}\widehat{\mathbf{V}} & \rightarrow\widehat{\mathbf{P}}=m_{0}\widehat{\gamma}\widehat{\mathbf{V}}+\widehat{\mathbf{A}},\\
\widehat{\lambda}_{x_{j}} & \rightarrow\widehat{\lambda}'_{x_{j}}=\widehat{\lambda}_{x_{j}}-\frac{\partial\widehat{A}_{i}}{\partial\widehat{X}_{j}}\widehat{\lambda}_{p_{i}},\\
\frac{1}{m_{0}}\widehat{\lambda}_{v_{i}} & \rightarrow\widehat{\lambda}_{p_{i}}=\frac{1}{m_{0}}\left[\widehat{\gamma}^{-1}\left(\widehat{\lambda}_{v_{i}}-V_{i}V_{k}\widehat{\lambda}_{v_{k}}\right)\right]_{S,}
\end{align}
can be obtained via the unitary operator
\begin{align*}
\widehat{C} & =\widehat{C}_{2}\widehat{C}_{1},\\
\widehat{C}_{1} & =\exp\left[\frac{i}{2}\left(\widehat{V}^{2}\,\widehat{\mathbf{V}}\cdot\lambda_{\mathbf{v}}\right)_{S}\right],\\
\widehat{C}_{2} & =\exp\left[i\mathbf{A}\cdot\lambda_{\mathbf{p}}\right].
\end{align*}

First, we have the immediate result $e^{i\widehat{C}}\widehat{\mathbf{R}}e^{-i\widehat{C}}=\widehat{\mathbf{R}}$
since $\widehat{C}$ does not depend on $\widehat{\lambda}_{x_{j}}$.
We can proceed for $\widehat{\lambda}_{\mathbf{r}}$ as follows: since
$\widehat{C}_{1}$ is not function of $\widehat{\mathbf{R}}$, we
have 

\begin{equation}
\widehat{C}\widehat{\lambda}_{x_{i}}\widehat{C}=\widehat{C}_{2}\widehat{\lambda}_{x_{i}}\widehat{C}_{2}^{-1}.
\end{equation}
Then, by a Baker-Campbell-Hausdorff expansion $e^{X}Ye^{-X}=\sum_{n=0}^{\infty}\frac{1}{n!}[X,Y]^{(n)}=Y+[X,Y]+\frac{1}{2}[X,[X,Y]]+\ldots$,
we have that

\begin{eqnarray}
e^{i\mathbf{\widehat{A}}\cdot\widehat{\lambda}_{\mathbf{p}}}\widehat{\lambda}_{x_{i}}e^{-i\mathbf{\widehat{A}}\cdot\widehat{\lambda}_{\mathbf{p}}} & = & \widehat{\lambda}_{x_{i}}+i[\mathbf{\widehat{A}}\cdot\widehat{\lambda}_{\mathbf{p}},\widehat{\lambda}_{x_{i}}]-[\mathbf{\widehat{A}}\cdot\widehat{\lambda}_{\mathbf{p}},[\mathbf{\widehat{A}}\cdot\widehat{\lambda}_{\mathbf{p}},\widehat{\lambda}_{x_{i}}]]+\ldots\nonumber \\
 & = & \widehat{\lambda}_{x_{i}}-\frac{\partial\widehat{A}_{j}}{\partial\widehat{X}_{i}}\widehat{\lambda}_{p_{j}}.
\end{eqnarray}

The procedure for $m_{0}\widehat{\mathbf{V}}$ and $\frac{1}{m_{0}}\widehat{\lambda}_{\mathbf{v}}$
is more involved as it entails showing equality between power series.
We start by considering the transformation of $m_{0}\widehat{\mathbf{V}}$.
Let us first define the kinematic momentum operator $\widehat{\varPi}=m_{0}\widehat{\gamma}\widehat{\mathbf{V}}.$
This operators obeys the following commutation relation
\[
\left[\widehat{P}_{j},\widehat{\lambda}_{p_{i}}\right]=\left[\widehat{\varPi}_{j},\widehat{\lambda}_{p_{i}}\right]=i\delta_{ij}.
\]
The transformation given by $\widehat{C}_{1}$ on $m_{0}\widehat{\mathbf{V}}$
is 
\begin{equation}
\widehat{C}_{1}\left(m_{0}\widehat{\mathbf{V}}\right)\widehat{C}_{1}^{-1}=m_{0}\widehat{\gamma}\widehat{V_{i}}=\widehat{\varPi}_{i}.\label{eq:C1V}
\end{equation}
We prove this affirmation by expanding both sides of (\ref{eq:C1V})
in power series. The left hand side can be expanded via the Baker-Campbell-Hausdorff
formula, whereas in the right hand side we can expand the Lorentz
factor $\widehat{\gamma}$ in a Maclaurin series as follows

\begin{align}
\widehat{C}_{1}\left(m_{0}\widehat{V_{i}}\right)\widehat{C}_{1}^{-1} & =m_{0}\widehat{V_{i}}+\frac{im_{0}}{2}[\left(\widehat{V}^{2}\,\widehat{\mathbf{V}}\cdot\lambda_{\mathbf{v}}\right)_{S},\widehat{V_{i}}]\ldots.\nonumber \\
 & =m_{0}\widehat{V_{i}}+\frac{m_{0}}{2}\widehat{V_{i}}\widehat{V}^{2}\ldots,\label{baker}\\
m_{0}\widehat{\gamma}\widehat{V_{i}} & =m_{0}\widehat{V_{i}}\sum_{n=0}^{\infty}\widehat{V}^{2n}\prod_{k=1}^{n}\left(\frac{2k-1}{2k}\right)=m_{0}\widehat{V_{i}}+\frac{m_{0}}{2}\widehat{V_{i}}\widehat{V}^{2}\ldots.\label{Lorentz}
\end{align}
We cans see that $\widehat{C}_{1}\left(m_{0}\widehat{V_{i}}\right)\widehat{C}_{1}^{-1}$
and $m_{0}\widehat{\gamma}\widehat{V_{i}}$ coincide at order $\widehat{V}^{2}$.
We can prove that they coincide at all orders by mathematical induction.
Let us assume the series agree up to the $nth$ term, this is, we
assume that

\begin{equation}
\frac{i^{n}}{2^{n}n!}[\left(\widehat{V}^{2}\,\widehat{\mathbf{V}}\cdot\lambda_{\mathbf{v}}\right)_{S},\widehat{V_{i}}]^{(n)}=\widehat{V_{i}}\widehat{V}^{2n}\prod_{k=1}^{n}\left(\frac{2k-1}{2k}\right).
\end{equation}
Then, the equality of the power series is proven by noting that they
also agree in the $n+1$ term. We have that

\begin{align}
\frac{i^{n+1}}{2^{n+1}(n+1)!}[\left(\widehat{V}^{2}\,\widehat{\mathbf{V}}\cdot\lambda_{\mathbf{v}}\right)_{S},\widehat{V_{i}}]{}^{(n+1)} & =\frac{i}{2(n+1)}[\left(\widehat{V}^{2}\,\widehat{\mathbf{V}}\cdot\lambda_{\mathbf{v}}\right)_{S},\widehat{V_{i}}\widehat{V}^{2n}\prod_{k=1}^{n}\left(\frac{2k-1}{2k}\right)]\nonumber \\
= & \widehat{V_{i}}\widehat{V}^{2n+1}\prod_{k=1}^{n+1}\left(\frac{2k-1}{2k}\right),
\end{align}
thus, Eq. (\ref{eq:C1V}) has veen proven. 

We now proceed to show that 
\begin{equation}
\widehat{C}_{2}\left(\widehat{\varPi}_{i}\right)\widehat{C}_{2}^{-1}=\widehat{\varPi}_{i}+\widehat{A}_{i}.\label{PPIA}
\end{equation}
This can be proven by noting that $\left[i\mathbf{A}\cdot\lambda_{\mathbf{p}},\widehat{\varPi}_{i}\right]=\widehat{A}_{i}$
and $\left[i\mathbf{A}\cdot\lambda_{\mathbf{p}},\left[i\mathbf{A}\cdot\lambda_{\mathbf{p}},\widehat{\varPi}_{i}\right]\right]=0$.
Equation (\ref{PPIA}) then follows by a Baker-Campbell-Hausdorff
expansion. In summary, we have shown that 
\begin{equation}
\widehat{C}\left(m_{0}\widehat{V_{i}}\right)\widehat{C}^{-1}=\widehat{C}_{2}\widehat{C}_{1}\left(m_{0}\widehat{V_{i}}\right)\widehat{C}_{1}^{-1}\widehat{C}_{2}^{-1}=\widehat{\varPi}_{i}+\widehat{A}_{i}=\widehat{P}_{i}.
\end{equation}

The transformation of $\widehat{\lambda}_{\mathbf{v}}$ involves,
again, a power series comparison. The claim is that 
\begin{equation}
\widehat{C}_{1}\left(\frac{1}{m_{0}}\widehat{\lambda}_{v_{i}}\right)\widehat{C}_{1}^{-1}=\frac{1}{m_{0}}\left[\widehat{\gamma}^{-1}\left(\widehat{\lambda}_{v_{i}}-V_{i}V_{k}\widehat{\lambda}_{v_{k}}\right)\right]_{S}.\label{LV}
\end{equation}
With the help of
\begin{equation}
\widehat{\gamma}^{-1}=1-\frac{1}{2}\widehat{V}^{2}-\frac{1}{8}\widehat{V}^{4}-\frac{1}{16}\widehat{V}^{6}+\ldots,
\end{equation}
we can write both sides of (\ref{LV}) as

\begin{align}
\widehat{C}_{1}\left(\frac{1}{m_{0}}\widehat{\lambda}_{v_{i}}\right)\widehat{C}_{1}^{-1} & =\frac{1}{m_{0}}\widehat{\lambda}_{v_{i}}+\frac{i}{2}[\left(\widehat{V}^{2}\,\widehat{\mathbf{V}}\cdot\lambda_{\mathbf{v}}\right)_{S},\frac{1}{m_{0}}\widehat{\lambda}_{v_{i}}]+\ldots,\nonumber \\
\frac{1}{m_{0}}\left[\widehat{\gamma}^{-1}\left(\widehat{\lambda}_{v_{i}}-V_{i}V_{k}\widehat{\lambda}_{v_{k}}\right)\right]_{S,} & =\frac{1}{m_{0}}\widehat{\lambda}_{v_{i}}-\frac{1}{m_{0}}\left(V_{i}V_{k}\widehat{\lambda}_{v_{k}}\right)_{S}\nonumber \\
 & -\frac{1}{2}\widehat{V}^{2}\left(\widehat{\lambda}_{v_{i}}-V_{i}V_{k}\widehat{\lambda}_{v_{k}}\right)\ldots.
\end{align}
This comparison is more elaborated than the previous one since the
brackets $[\left(\widehat{V}^{2}\,\widehat{\mathbf{V}}\cdot\lambda_{\mathbf{v}}\right)_{S},\frac{1}{m_{0}}\widehat{\lambda}_{v_{i}}]^{(n)}$
produce terms of different order in $\widehat{V}^{2}$. However, it
can be checked that the first terms of both series agree, and the
equality of the entire series can be established, again, by mathematical
induction. Since $\widehat{C}_{2}\widehat{\lambda}_{p_{i}}\widehat{C}_{2}^{-1}=\widehat{\lambda}_{p_{i}}$,
we finally obtain 

\begin{equation}
\widehat{C}\left(\frac{1}{m_{0}}\widehat{\lambda}_{v_{i}}\right)\widehat{C}^{-1}=\frac{1}{m_{0}}\left[\widehat{\gamma}^{-1}\left(\widehat{\lambda}_{v_{i}}-V_{i}V_{k}\widehat{\lambda}_{v_{k}}\right)\right]_{S,}=\widehat{\lambda}_{p_{i}}.
\end{equation}


\begin{thebibliography}{10}
\bibitem{weinberg}S. Weinberg, ``The Quantum Theory of Fields''
(Cambridge University Press, Cambridge, U.K., 1995), Vol. 1.

\bibitem{ohnuki}Y. Ohnuki, ``Unitary Representations of the Poincar\'e
Group and Relativistic Wave Equations'' (World Scientific, Singapore,
1988)

\bibitem{prosperi}M. Pauri and G. M. Prosperi, J. Math. Phys. $\mathbf{16}$,
1503 (1975).

\bibitem{sudarshan}E. C. G. Sudarshan and N. Mukunda, ``Classical
Dynamics: A Modern Perspective'' (Wiley, New York, 1974).

\bibitem{cla1}B. O. Koopman, Proc. Natl. Acad. Sci. U.S.A. $\mathbf{17}$,
315 (1931).

\bibitem{cla 2}J. von Neumann, Ann. Math. $\mathbf{33}$, 587 (1932);
ibid. $\mathbf{33}$, 789 (1932).

\bibitem{cla 3}D. Mauro. ``Topics in Koopman-von Neumann Theory''
PhD thesis, Universit\`{a} degli Studi di Trieste, (2002), arXiv:quantph/0301172
{[}quant-ph{]}.

\bibitem{cla 5}D. Bondar, et al, Phys. Rev. Lett. $\mathbf{109}$,
190403 (2012).

\bibitem{cla 6}U. Klein, Quantum Stud.: Math. Found. $\mathbf{5}$,
219 (2018) .

\bibitem{Cla9}J. Wilkie and P. Brumer, Phys. Rev. A $\mathbf{55}$,
27 (1997); Phys. Rev. A $\mathbf{55}$, 43 (1997).

\bibitem{Cla 7}D. Mauro, Int. J. Mod. Phys. A $\mathbf{17}$, 1301
(2002).

\bibitem{Cla 8}E. Gozzi, and D. Mauro, Int. J. Mod. Phys. A $\mathbf{19}$,
1475 (2004).

\bibitem{minimal coupling}E. Gozzi and D. Mauro, Ann. Phys (NY). $\mathbf{296}$
(2002) 152-186.

\bibitem{cla9}P. Brumer and J. Gong, Phys. Rev. A. $\mathbf{73}$,
052109 (2006).

\bibitem{cla10}S. Katagiri, Prog. Theor. Exp. Phys. $\mathbf{6}$, 063A02 (2020).

\bibitem{cla11} P. Morgan, Ann. Phys (NY). $\mathbf{414}$ (2020) 168090.
\bibitem{cla12}A. Sen, Z. Silagadze, Int. J. Theor. Phys. $\mathbf{59}$ ,2187-2190 (2020).
\bibitem{cla13}I. Ramos-Prieto, A.R Urz\'{u}la-Pineda, F. Soto-Eguibar, and H.M Moya-Cessa. Sci. Rep. $\mathbf{8}$ , 8401
(2018).
\bibitem{cla14} I. Joseph, Phys. Rev. Research.  $\mathbf{2}$ , 043102  (2020).
\bibitem{cla15} O.I. Chashchina, A. Sen and Z. K. Silagadze, Int. J. Mod. Phys. D $\mathbf{29}$ (2020), 2050070.
\bibitem{cla16} A. Sen, S. Dhasmana, Z. K. Silagadze, Ann. Phys. $\mathbf{414}$ (2020), 168302.

\bibitem{cla17} C. Barcel\'{o}, R. Carballo-Rubio, L. J. Garay, and R. G\'{o}mez-
Escalante, Phys. Rev. A $\mathbf{86}$, 042120 (2012).

\bibitem{cla18} A. Peres and D. R. Terno, Phys. Rev. A $\mathbf{63}$, 022101 (2001).
\bibitem{cla19}D. I. Bondar, F. Gay-Balmaz , C. Tronci. Proc. R.
Soc. A $\mathbf{475}$, 20180879 (2019).
\bibitem{cla20} F. Gay-Balmaz , C. Tronci , Nonlinearity $\mathbf{33}$ (2020) 5383-5424.

\bibitem{galilei}A. D. Berm\'{u}dez Manjarres, M. Nowakowski, and D.
Batic, Ann. Phys (NY). $\mathbf{416}$, 168157 (2020).

\bibitem{galilei2}A. D. Berm\'{u}dez Manjarres and N. Mar\'{i}n-Medina, Phys.
Rev. A. $\mathbf{102}$, 042221 (2020).

\bibitem{sudarshan-2}E. C. G. Sudarshan, Pramana $\mathbf{6}$, 117
(1976).

\bibitem{arnold}V. I. Arnold and A. Avez, ``Ergodic Problems of
Classical Mechanics'' (Benjamin, New York, 1968).

\bibitem{bopp}D. I. Bondar, R. Cabrera, D. V. Zhdanov, and H. A. Rabitz, Phys. Rev. A. $\mathbf{88}$, 052108 (2013)

\bibitem{ballentine}L. E. Ballentine, ``Quantum Mechanics: A Modern
Development'', pp 63-86 (World Scientific Publishing, Singapore;
1998).

\bibitem{jordan}T. F. Jordan, ``Linear Operators for Quantum Mechanics'',
(John Wiley \& Sons, New York; 1969).

\bibitem{barut}A.O. Barut, ``Electrodynamics and classical theory
of fields and particles'', pp 54-57 (Dover Publications, Mineola,
2010).

\bibitem{co}L. J. Na\dj \dj er\dj , M. D. Davidovi\'{c}, and D. M.
Davidovi\'{c}, Am. J. Phys. $\mathbf{82}$, 1083 (2014).

\bibitem{QCT}A. Anderson, Ann. Phys. $\mathbf{232}$ (1994) 292-331.

\bibitem{Vjordan}T. F. Jordan, J. Math. Phys. $\mathbf{18}$, 608
(1977).

\bibitem{newton-wigner}T. D. Newton and E. P. Wigner, Rev. Mod. Phys.
$\mathbf{21}$, 400 (1949).

\bibitem{newton-wigner2}A. H. Monahan and M. McMillan, Phys. Rev.
A $\mathbf{56}$, 2563 (1997).

\bibitem{clanewton}P. K. Schwartz and D. Giulini, Int. J. Methods
mod. Phys. $\mathbf{17}$, 12, 2050176 (2020).
\end{thebibliography}
\end{document}